\documentclass[final,5p,times,twocolumn]{elsarticle}   

\usepackage{amssymb}

\usepackage{lineno}

\journal{Nuclear Instruments and Methods A}

\usepackage{amssymb}

\begin{document}

\begin{frontmatter}


\title{     Identification of medium mass (A=60--80) ejectiles
            from 15 MeV/nucleon peripheral heavy-ion collisions with the MAGNEX 
            large-acceptance spectrometer   }



\author[a]{ G. A. Souliotis\corref{cor1} }         
\author[a]{S. Koulouris} 
\author[b,c]{F. Cappuzzello}  
\author[c]{D. Carbone} 
\author[d]{A. Pakou}
\author[c]{C. Agodi}     
\author[b,c]{G. Brischetto}  
\author[c]{S. Calabrese}  
\author[c]{M. Cavallaro} 
\author[b,c]{I. Ciraldo} 
\author[e]{J. Klimo}     
\author[c]{O. Sgouros}  
\author[b,c]{V. Soukeras}  
\author[b,c]{A. Spatafora} 
\author[c]{D. Torresi}  
\author[f]{M. Veselsky}  

\address[a]{Laboratory of Physical Chemistry, Department of Chemistry,
                 National and Kapodistrian University of Athens, Athens, Greece}

\address[b]{Dipartimento di Fisica e Astronomia "Ettore Majorana", 
                Universita di Catania, Italy}
                   
\address[c]{Laboratori Nazionali del Sud, INFN, Italy} 
           
\address[d]{Department of Physics and HINP, The University of Ioannina, Ioannina, Greece}

\address[e]{Institute of Physics, Slovak Academy of Sciences, Bratislava, Slovakia}

\address[f]{Institute of Experimental and Applied Physics, Czech Technical University, 
            Prague, Czech Republic}


\cortext[cor1]{Corresponding author, e-mail: soulioti@chem.uoa.gr}



\begin{abstract}

An approach to identify medium-mass ejectiles from peripheral heavy-ion reactions 
in the energy region of 15 MeV/nucleon  is developed
for data obtained with a large acceptance magnetic spectrometer. This spectrometer is equipped
with a focal plane multidetector, providing position, angle, energy loss and residual energy 
of the ions along with measurement of the time-of-flight. 
Ion trajectory reconstruction is performed at high order and ion mass is obtained with a
resolution of better than 1/150. For the unambiguous particle identification however, the
reconstruction of both the atomic number Z and the ionic charge q of the ions is critical and
it is suggested, within this work, to be  performed prior to mass identification. 
The new proposed method was successfully applied to  MAGNEX spectrometer data, for identifying
neutron-rich ejectiles related to multinucleon transfer generated in the $^{70}$Zn+$^{64}$Ni collision at 15 MeV/nucleon.  
This approach opens up the possibility of employing heavy-ion reactions with medium-mass beams
below the Fermi energy (i.e., in the region 15--25 MeV/nucleon) in conjunction with 
large acceptance ray tracing spectrometers, first, to study the mechanism(s) of nucleon transfer
in these reactions and, second, to produce and study very neutron-rich or even new nuclides 
in previously unexplored regions of the nuclear landscape.

\end{abstract}

\end{frontmatter}





\section{Introduction }

The study of exotic nuclei  toward the neutron drip-line has recently received strong attention
by the nuclear physics community  (see, e.g., \cite{ndrip0,ndrip1,ndrip2}). 
Toward this goal, the efficient production of very neutron-rich nuclides constitutes a central 
issue in current and upcoming facilities around the world
(see, e.g., \cite{FRIB1,FRIB2,GANIL,GSI,RIBF,ARGONNE,LNS,RISP}).           
Traditional routes to produce neutron-rich nuclides  are spallation, fission and projectile 
fragmentation \cite{RIB-2013}.  Surpassing the limits of the above approaches constitutes a 
vigorous endeavor of nuclear scientists. 
Toward this end, to access nuclides with high neutron-excess, besides proton stripping 
from the projectile in the context of fragmentation, it is necessary to capture neutrons 
from the target. Such transfer takes place in peripheral nucleon-exchange reactions at beam energies
from the Coulomb barrier \cite{ndrip1,Volkov-1978,Dasso-1994,Zagrebaev-2008,Corradi-2009,Corradi-2013} 
to the Fermi energy (15--35 MeV/nucleon) \cite{GS-PLB02,GS-PRL03}.

An extensive effort to exploit multinucleon transfer reactions near the Coulomb barrier
combined with large acceptance spectrometers to produce and study neutron rich 
isotopes has been devoted in recent years.  The wide angular and energy distribution 
of the products from such reactions necessitates the use of new-generation spectrometers
having large angular and momentum acceptance. Recent  representative reactions are 
$^{40}$Ca+$^{96}$Zr, $^{90}$Zr+$^{208}$Pb \cite{Szilner-2007},
$^{40}$Ar+$^{208}$Pb \cite{Mijatovic-2016a,Mijatovic-2016b},
$^{136}$Xe+$^{238}$U \cite{Vogt-2015} and
$^{197}$Au+$^{130}$Te \cite{Galtarossa-2018}  studied with the PRISMA spectrometer \cite{Szilner-2007,Montanari-2011} at INFN-LNL, as well as,
$^{136}$Xe+$^{198}$Pt \cite{Watanabe-2015} and $^{238}$U+$^{12}$C \cite{Caammano-2013}
investigated with the VAMOS spectrometer \cite{Rejmund-2011,Pullanhiotan-2008,Savajols-2003}
at GANIL.

As a complementary approach to access neutron rich nuclei, and guided by experience in reaction
dynamics near and below the Fermi energy, we explored peripheral heavy-ion reactions in the 
energy range 15--25 MeV/nucleon.
Our initial experimental studies of projectile fragments from 25 MeV/nucleon reactions of $^{86}$Kr 
on $^{64}$Ni \cite{GS-PLB02} and $^{124}$Sn \cite{GS-PRL03} indicated substantial production of 
neutron rich fragments especially near the projectile.
Subsequently, motivated by developments in several facilities offering either intense primary beams
\cite{GANIL,ARGONNE,LNS} at this energy range, or re-accelerated rare isotope beams
\cite{FRIB2,GANIL,ARGONNE,RISP}, we continued our studies 
at 15 MeV/nucleon  \cite{GS-PRC11,Fountas-2014,Papageorgiou-2018,Palli-2021}, where we observed further enhancement in the cross sections of neutron rich isotopes very close to the projectile.

Our experimental studies were based on the use of a medium-acceptance spectrometer,  i.e. the MARS
recoil separator \cite{MARS,GS-NIM03,GS-NIM08} at the Cyclotron Institute of Texas A\&M University.
However, our work with the heavier  
targets $^{124}$Sn and  $^{208}$Pb \cite{GS-PRL03,GS-NIM03} indicated the limitations of a 
typical forward separator to access the very neutron-rich products that  are concentrated at angles 
near the grazing angle, as in reactions near and above the Coulomb barrier.
Furthermore, our recent calculations with the heavy target $^{238}$U \cite{Papageorgiou-2018}
indicated the advantage of this target to access the most neutron-rich nuclides.
Thus, for efficient collection of these fragments, the use of a large acceptance separator 
is indispensable, as in the case of reactions near the Coulomb barrier.  
Of course, in the energy region of 15--25 MeV/nucleon, the velocities of the ejectiles are higher and the angular distributions are narrower, compared with the Coulomb barrier reactions, thus, favoring efficient collection and identification.
 
For these reasons, we initiated a project to produce and identify  projectile-like
fragments in the MAGNEX large-acceptance spectrometer at INFN-LNS  from the reaction of 
a $^{70}$Zn (15 MeV/nucleon) beam on a $^{64}$Ni target. 
We note that MAGNEX has been designed as a tool for charged particle spectroscopy, able to guarantee good energy resolution, angular resolution and also able to measure accurate absolute cross sections for the selected ions of interest. 
With the present effort, we investigated its application to reactions involving medium-mass 
heavy ions, where the Z resolution appears limited. Thus, we had to develope a detailed 
approach to reconstruct the absolute value of the atomic number Z of the ejectiles, 
along with their ionic charge states, based on carefully calibrated energy loss, energy and 
time-of-flight measurements.
In this article, we describe the adopted identification techniques and we outline 
the physics possibilities.
The structure of the paper is as follows.
In section 2, the experimental setup and measurements are outlined. In section 3, the data analysis
is described with emphasis on the identification of heavy projectile-like fragments. 
Finally a discussion and conclusions are  given in section 4.


\section{Outline of the experimental setup and measurements }

The experiment was performed at the MAGNEX facility of Istituto Nazionale di Fisica Nucleare,
Laboratori Nazionali del Sud (INFN-LNS) in Catania, Italy.
A beam of $^{70}$Zn$^{15+}$  at 15 MeV/nucleon delivered by the K800 superconducting cyclotron
bombarded a 1.18 mg/cm$^{2}$ $^{64}$Ni foil    
at the optical object point of the MAGNEX large acceptance spectrometer
\cite{Cunsolo-2002a,Cunsolo-2002b,Cappuzzello-2016}.
The optical axis of the spectrometer was set at an angle of 9.0$^{o}$ spanning the  interval 
4.0$^{o}$--15.0$^{o}$. 
 
The unreacted beam was collected in an electron suppressed Faraday cup inside the target chamber.
The ejectiles emerging from the target passed through a 6 $\mu$m Mylar stripper foil and then were 
momentum analysed by the MAGNEX spectrometer \cite{Cappuzzello-2010,Cappuzzello-2011} and 
detected by its focal plane detector (FPD)  \cite{Cavallaro-2012,Torresi-2021}.


The MAGNEX spectrometer consists of a large aperture vertically focusing quadrupole magnet 
and a horizontally bending dipole magnet. 
MAGNEX allows the identification of heavy ions with resolutions  of $\Delta$A/A $\sim $1/160 
in mass,  $\Delta\theta$ $\sim$ 0.2$^{o}$ in angle  and $\Delta$E/E $\sim$ 1/1000 in energy
within a large solid angle of $\Delta \Omega$ $\sim$ 50 msr and momentum range 
of -14\% $\leq$  $\Delta$p/p  $\leq$ +10\%. 
This performance results from the implementation of trajectory reconstruction, based on differential
algebraic techniques, which allows solving the equation of motion of each detected ion to 10th order.

The focal plane detector FPD \cite{Torresi-2021} is a large 
(1360$\times$200$\times$90 mm$^{3}$ active volume)
gas-filled hybrid detector with a wall of 60 large-area silicon detectors 
(50$\times$70$\times$0.5 mm$^{3}$) arranged in three rows at the end. 
[Schematic drawings of FPD are presented in Figs. 1 and 2 of \cite{Torresi-2021}.]
The detector had a 6 $\mu$m Mylar foil at the entrance window and was operated with isobutane gas at a pressure of 40 mbar. 
The detector provided the horizontal and vertical coordinates (x$_{f}$,y$_{f}$) of each
incident particle at six sequential planes crossed by the ion trajectory, thus
accurately  determining the  angles $\theta_f$, $\phi_f$ of the ion's trajectory \cite{Torresi-2021}.
It also provides the energy loss in the gas and the residual energy deposited in
the silicon detectors. 
Moreover, the signals of the silicon detectors gave the start for the time of flight
(TOF) measurement of the particles through the spectrometer, while the stop signal was 
provided by the radiofrequency of the cyclotron. 
This setup provided a modest TOF resolution of $\sim$3 ns,
limited by the cyclotron RF timing. (The possibility of improving this 
to $\sim$1 ns will be pursued for future runs.)
We mention that the expected TOF of ions with the beam velocity along 
the central trajectory is 114 ns up to the FPD. 
In this exploratory run, we did not employ a timing detector (e.g. a micro-channel plate electron amplifier) after the target, an option that we will explore for future runs.


In the present experiment, aiming mainly at the feasibility of ejectile identification, 
only about one-half of the active area of FPD was used
(the other part was covered with an aluminum screen)
in order to avoid radiation damage of the silicon detectors and 
high dead times due to limitations in the data acquisition system.
These experimental restrictions  will be circumvented in the future in view 
of the upgrading of the MAGNEX facility \cite{Agodi-2021}.
Thus, in a full acceptance run, the whole area of the FPD can be exposed to the flux 
of the reaction products, especially for the investigation of very suppressed reaction 
channels.
Moreover, a set of vertical slits before the quadrupole restricted the vertical angular 
acceptance of MAGNEX in the range -0.8$^{o}$ to 0.8$^{o}$. 
It should be noted that opening the vertical acceptance toward 
its full range (-7.1$^{o}$ to 7.1$^{o}$) is not expected to affect significantly 
the detector resolution and the overall spectrometer performance. 
This is due to the trajectory  reconstruction procedure that provides compensation 
for the ensuing large aberrations. 

The restriction of the vertical acceptance and the active area of the FPD resulted 
in the use of only seven of the silicon detectors belonging to the middle row. 
The magnetic rigidity  was set in order to transport the $^{70}$Zn$^{29+}$ 
ejectiles and bring them to the center of FPD.
\begin{figure}[ht]                    
\centering
\includegraphics[width=0.45\textwidth,keepaspectratio=true]{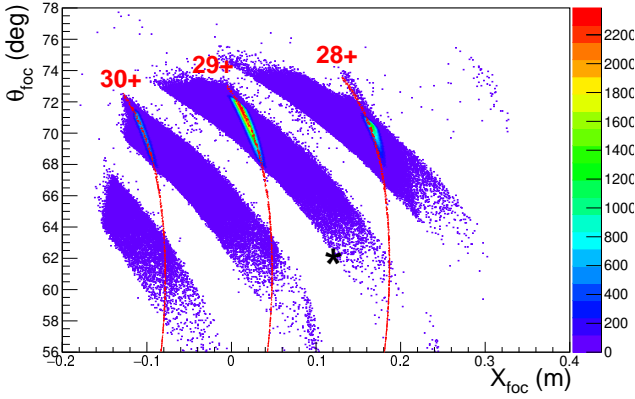}   
\caption{ (Color online)
Plot of the horizontal angle ($\theta_{foc}$, in degrees) versus the horizontal position 
(X$_{foc}$, in meters) measured at the MAGNEX focal plane for ejectiles from 
the interaction of a $^{70}$Zn beam (15 MeV/nucleon) with a $^{64}$Ni target
distributed on four of the FPD silicon detectors.
The open (blank) areas correspond to three other silicon detectors (whose data
are omitted by software gates) adjacent to the ones whose data are shown.
The asterisk designates the Si detector with which the correlations displayed in 
Figs. 2--6 are generated.
The red lines are simulations for elastically scattered $^{70}$Zn ejectiles in the 
charge states ${30+}$, ${29+}$ and ${28+}$, from left to right, respectively.
}
\label{fig01}
\end{figure}
A typical plot of the horizontal angle ($\theta_{foc}$, in degrees) versus the horizontal position
(X$_{foc}$, in meters) measured at the FPD is presented in Fig. 1. 
Unidentified experimental data of ejectiles from the $^{70}$Zn (15 MeV/nucleon) + $^{64}$Ni reaction,  
distributed on four of the FPD silicon detectors, are shown in the figure.  
The blank bands correspond to data of 3 other silicon detectors, eliminated by software gates, 
for better clarity of the presentation.
Moreover, the red lines correspond to  simulated data for elastically scattered $^{70}$Zn 
ejectiles in the ionic charge states 30+, 29+ and 28+, from left to right, respectively, 
being in their ground state.  The agreement of the simulations with the experimental  
data is very good. 

\begin{figure}[ht]                     
\centering
\includegraphics[width=0.45\textwidth,keepaspectratio=true]{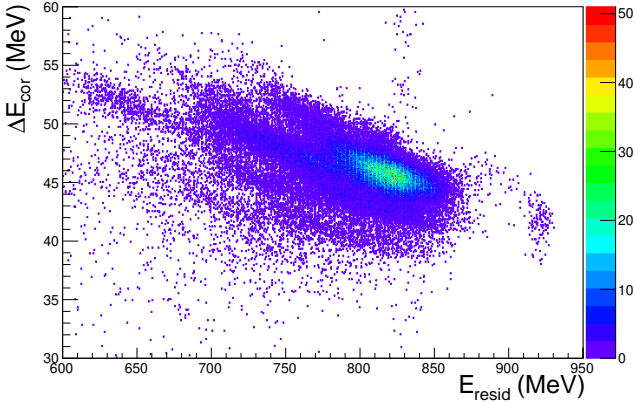} 
\caption{ (Color online) 
Representative $\Delta$E$_{cor}$ vs E$_{resid}$ plot 
of ejectiles from  the $^{70}$Zn (15 MeV/nucleon) + $^{64}$Ni
reaction corresponding to one silicon detector, for the same
data as in Fig. 1. 
This detector  is represented by the 3rd band from the left in Fig. 1
and is marked with and asterisk.
The elastic events of $^{70}$Zn$^{29+}$ have been gated out.
}
\label{fig02}
\end{figure}

\section{ Particle identification procedure}

The ejectiles of interest from the reaction $^{70}$Zn + $^{64}$Ni at 15 MeV/nucleon, 
studied in this work have atomic numbers  24$\leq$Z$\leq$32 and  mass numbers  60$<$A$<$80.

We first attempted to obtain the  atomic number of the ejectiles employing the standard
$\Delta$E--E correlation. 
A two-dimensional $\Delta$E--E plot is shown in Fig. 2 for one silicon detector.
This detector corresponds to the events of the 3$^{rd}$ band from the left in Fig. 1
and is marked with an asterisk.
Additionally, in Fig. 2, the elastic events of $^{70}$Zn$^{29+}$ have been eliminated
by a software gate set  on the $\theta_{foc}$ vs $X_{foc}$
correlation (Fig. 1) for the 29+ charge state peak (middle intense peak).
This plot involves the residual energy measured by the silicon detector (E$_{resid}$) 
and the total energy loss in the gas section of FPD  corrected for 
path length differences depending on the angle of incidence. 
This quantity is named  $\Delta$E$_{cor}$ \cite{Torresi-2021} (see also below).
In the $\Delta$E$_{cor}$ -- E$_{resid}$ spectrum, despite the appearance of bands,
presumably belonging to successive Z and having various charge states q, 
we are not able to proceed to a clear Z identification.
We should underline here that for reactions with  lighter projectiles, as $^{20}$Ne at 15 MeV/nucleon,
the $\Delta$E--E correlation is typically adequate for Z identification (e.g. \cite{Cavallaro-2020a}).

With the current design and operating conditions of the gas section of FPD 
(active length of 8 cm and gas pressure of 40 mbar),
for the medium mass ejectiles of 15 MeV/nucleon energy, 
the fraction of the energy loss with respect to the ion total energy is rather low,
$\sim$5--10\% (Fig. 2). 
This should be compared to the optimum empirical value of  $\sim$20--30\% 
(or higher), as reported in recent studies at Coulomb barrier energies 
(see, e.g., \cite{Caammano-2013} Fig. 4,   \cite{Rejmund-2011} Fig. 7 
and \cite{Szilner-2007} Fig. 1) in which clear Z identification was achieved
for medium-mass heavy ions.
Therefore, for the analysis of the present data and, more generally,  data where the 
standard $\Delta$E--E correlation is not adequate,  we can proceed as described 
in the following.

We developed a detailed procedure to reconstruct the atomic number Z of the ejectiles
based on the measured quantities $\Delta$E$_{cor}$, E$_{resid}$ 
and TOF, after absolute calibration of these quantities outlined below.
The energy loss in the FPD and the residual energy in the Si detectors were calibrated 
using the elastic peaks. 
In the calibration procedure, we took advantage of the fact that the elastically 
scattered $^{70}$Zn ions enter the FPD having a broad range of angles
(as indicated in Fig. 1), thus providing  calibration points 
of $\Delta$E$_{tot}$ and E$_{resid}$ for successive $\theta_{foc}$ angle windows.
The necessary energy loss calculations were based on the stopping power tables 
of Hubert et al. \cite{Hubert-1989}.
Furthermore,
for each event, the total energy loss $\Delta$E$_{tot}$ in the FPD was corrected for the angle
of incidence at the focal plane, giving the corrected (or reduced) energy loss $\Delta$E$_{cor}$
corresponding to the angle of incidence $\theta_{foc}$=59.2$^o$ of the central trajectory. 
Also the total kinetic energy E$_{tot}$  of the ions reaching FPD was obtained as the sum
of the measured energy loss $\Delta$E$_{tot}$ in the gas section of FPD and  the residual 
energy E$_{resid}$ 
in the silicon detector, including a calculated correction for the energy loss in the entrance 
window $\Delta$E$_{w}$ :

$$ E_{tot} = \Delta E_{w} + \Delta E_{tot} +  E_{resid} $$

Finally, the TOF was calibrated using the known velocity of the elastic events 
and their path lengths up to the Si detector wall that were obtained from the trajectory 
reconstruction procedure. Then, the velocity of each ion was determined from the 
calibrated TOF and the length of its trajectory with a resolution of $\sim$ 2.5\%.

Using the above calibrated quantities, and guided by the relation
$ Z \propto \upsilon \sqrt{\Delta E} $ from Bethe-Bloch equation,
we developed two approaches,  I and II, based on \cite{Souliotis-1998},
for the Z reconstruction that are presented  below.


In approach I, taking into account the velocity and $\Delta$E$_{cor}$, Z was 
reconstructed using the expression: 

\begin{eqnarray}
Z_{I} = a_0(\upsilon ) + a_1(\upsilon )\,\upsilon \sqrt{\Delta E_{cor}} + 
a_2(\upsilon)(\upsilon \sqrt{\Delta E_{cor}})^2   \label{Z_I_eq}
\end{eqnarray}

In order to determine the functions $a_0(\upsilon)$, $a_1(\upsilon)$ and 
$a_2(\upsilon)$ in the velocity range of interest, we used the energy-loss data of
Hubert et al. \cite{Hubert-1989}  to determine the coefficients of Eq. (\ref{Z_I_eq})  
for the  Z range 6--36 and in the energy range 8--18 MeV/nucleon by applying a
least-squares fitting procedure at each energy in steps of 0.5 MeV/nucleon.
The values of each coefficient at the various energies were then
fitted with polynomial functions of velocity. 

In approach II,  based on the velocity and both $\Delta$E$_{cor}$ and  E$_{tot}$, 
Z is reconstructed according to the expression: 

\begin{equation}
Z_{II} = b_0(\upsilon) + b_1(\upsilon)\, \sqrt{ \Delta E_{cor}\, E_{tot} } + 
b_2(\upsilon) (\sqrt{\Delta E_{cor} E_{tot} })^2   \label{Z_II_eq}
\end{equation}
 
As in approach I, the functions $b_0(\upsilon)$, $b_1(\upsilon)$ and 
$b_2(\upsilon)$ were determined in the above Z and velocity range 
of interest. 

 
We furthermore proceeded with the reconstruction of the ionic charge q 
of the ions based on the equation:
 
\begin{eqnarray}
 q & = & \frac{2 E_{tot}}{B\rho} \frac{TOF}{L}
\end{eqnarray}

in which $E_{tot}$ is the ion kinetic energy, B$\rho$ the magnetic rigidity, 
TOF the time of flight and L the length of the ion's trajectory, as discussed above.
The magnetic rigidity was obtained as: 
\begin{eqnarray}
 B\rho = B\rho_{0} ( 1 + \delta )
\end{eqnarray}
where the fractional deviation  $ \delta $  from the magnetic rigidity $ B\rho_{0} $ 
of the central trajectory was obtained from the trajectory reconstruction
procedure.

\begin{figure}[ht]                 
\centering
\includegraphics[width=0.45\textwidth,keepaspectratio=true]{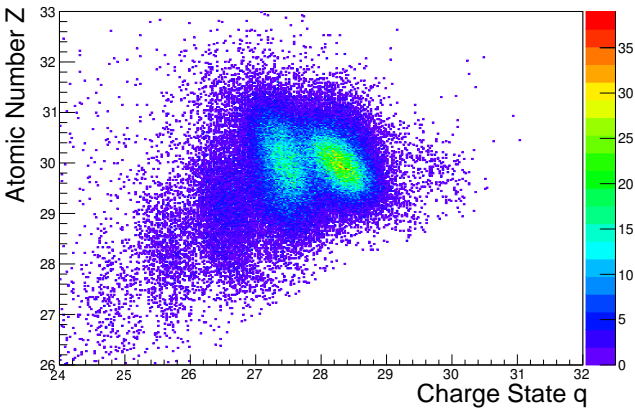}
\caption{ (Color online)   
Reconstructed Z$_{I}$ vs charge state q correlation  
of ejectiles from  the $^{70}$Zn (15 MeV/nucleon) + $^{64}$Ni
reaction corresponding to the same silicon detector  and conditions as in Fig. 2.
}
\label{fig03}
\end{figure}


\begin{figure}[ht]               
\centering
\includegraphics[width=0.45\textwidth,keepaspectratio=true]{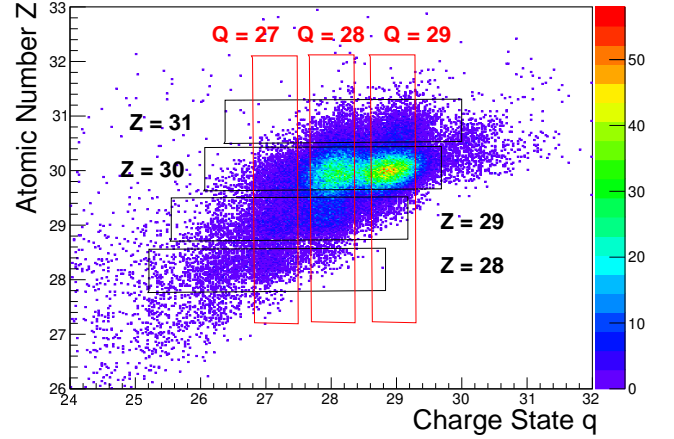}
\caption{ (Color online)   
Reconstructed Z$_{II}$ vs charge state q correlation  
of ejectiles from  the $^{70}$Zn (15 MeV/nucleon) + $^{64}$Ni
reaction corresponding to the same silicon detector and conditions as in Figs. 2 and 3.
Graphical contours are shown on each band corresponding to the atomic numbers 
Z (horizontal bands) and the ionic charge states q (vertical bands) 
of the ejectiles.
}
\label{fig04}
\end{figure}

Having determined the Z (with approaches I and II, denoted Z$_{I}$ and  Z$_{II}$, respectively), 
we proceed to obtain a correlation between Z and q as shown in Figs. 3 and 4.
The correlation of Z$_{I}$ with q (Fig. 3) does not appear to provide a Z separation, 
whereas that of Z$_{II}$ with q (Fig. 4) shows a rather adequate Z separation. 
At this point, we note that the Z$_{I}$ reconstruction based essentially on 
TOF and $\Delta E_{tot}$ shows inadequate resolution due to the  limited resolution
of the TOF measurement.
However,   the  Z$_{II}$ reconstruction based essentially on the measured $\Delta E_{cor}$  
and $E_{resid}$ (employing the TOF measurement only in the velocity-dependent coefficients) 
reaches a rather adequate resolution. 
From 1D projections of the Z--q correlation of Fig. 4, we can estimate the achieved
resolutions (FWHM) of Z and q to be $\sim$0.8 and $\sim$0.7 units, respectively, for ejectiles near 
the projectile. 



For the mass determination, we first proceed by setting gates on Z and q 
based on the Z reconstruction with approach II, Eq. (2), as seen in Fig. 4.
Subsequently, for given Z and q, we adopt the identification technique for large
acceptance spectrometers  introduced in Ref. \cite{Cappuzzello-2010}.
This technique is based on the relationship between  the total kinetic energy 
of the ions and their magnetic rigidity B$\rho$,  determined as discussed 
above in relation to Eq. (4): 

\begin{eqnarray}
B\rho  = & \frac{\sqrt{m}}{q} \sqrt{2 E_{tot}}
\end{eqnarray}

This relationship  depends on the ratio $\sqrt{m}$/q.
Thus, in a B$\rho$  versus E$_{tot}$ representation, the particles are distributed 
on different bands according to their  $\sqrt{m}$/q.
Furthermore, since we have selected a given q, the bands should correspond to the 
various masses for the specific Z and q.

\begin{figure}[ht]                         
\centering
\includegraphics[width=0.45\textwidth,keepaspectratio=true]{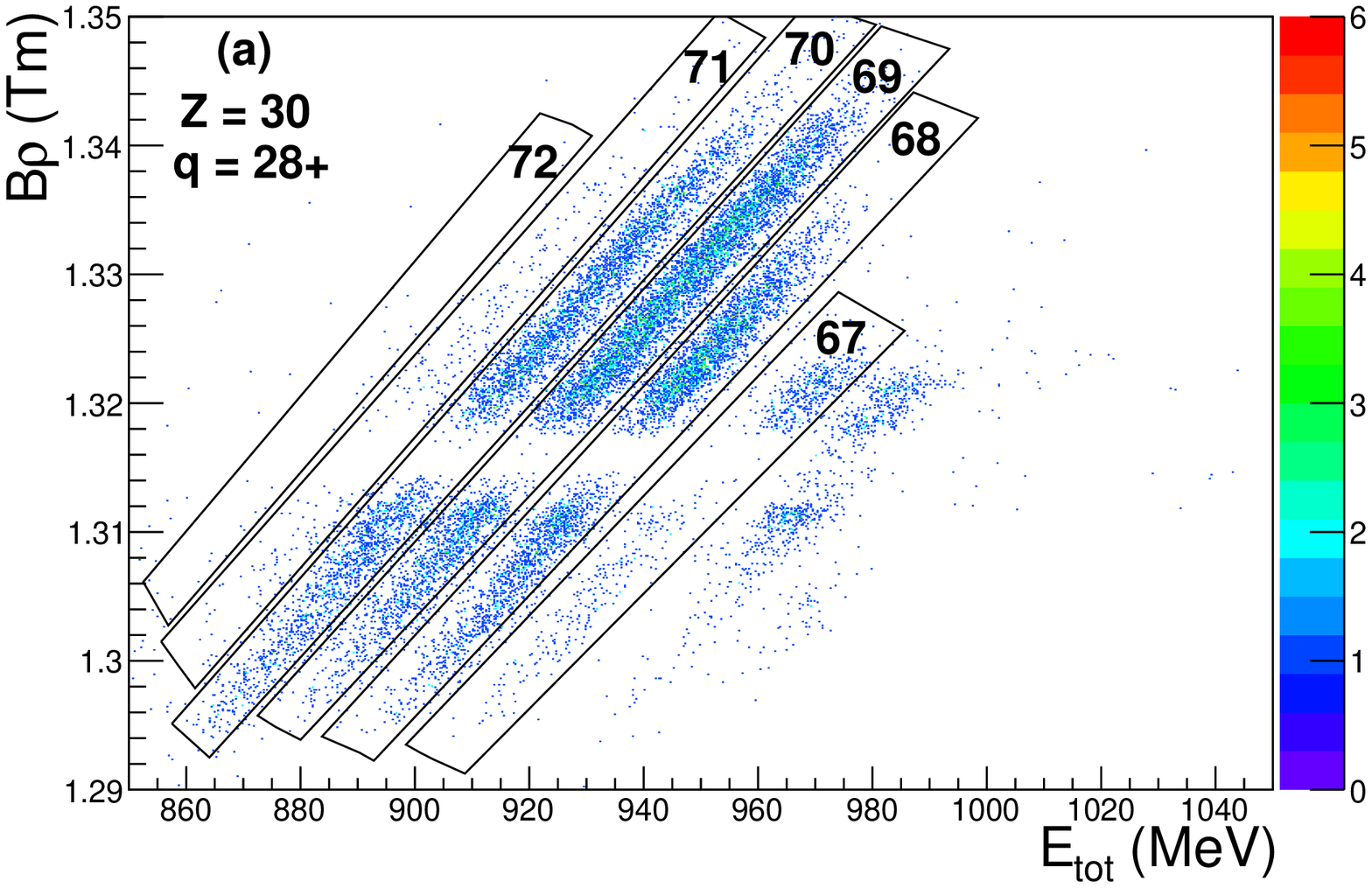}
\includegraphics[width=0.45\textwidth,keepaspectratio=true]{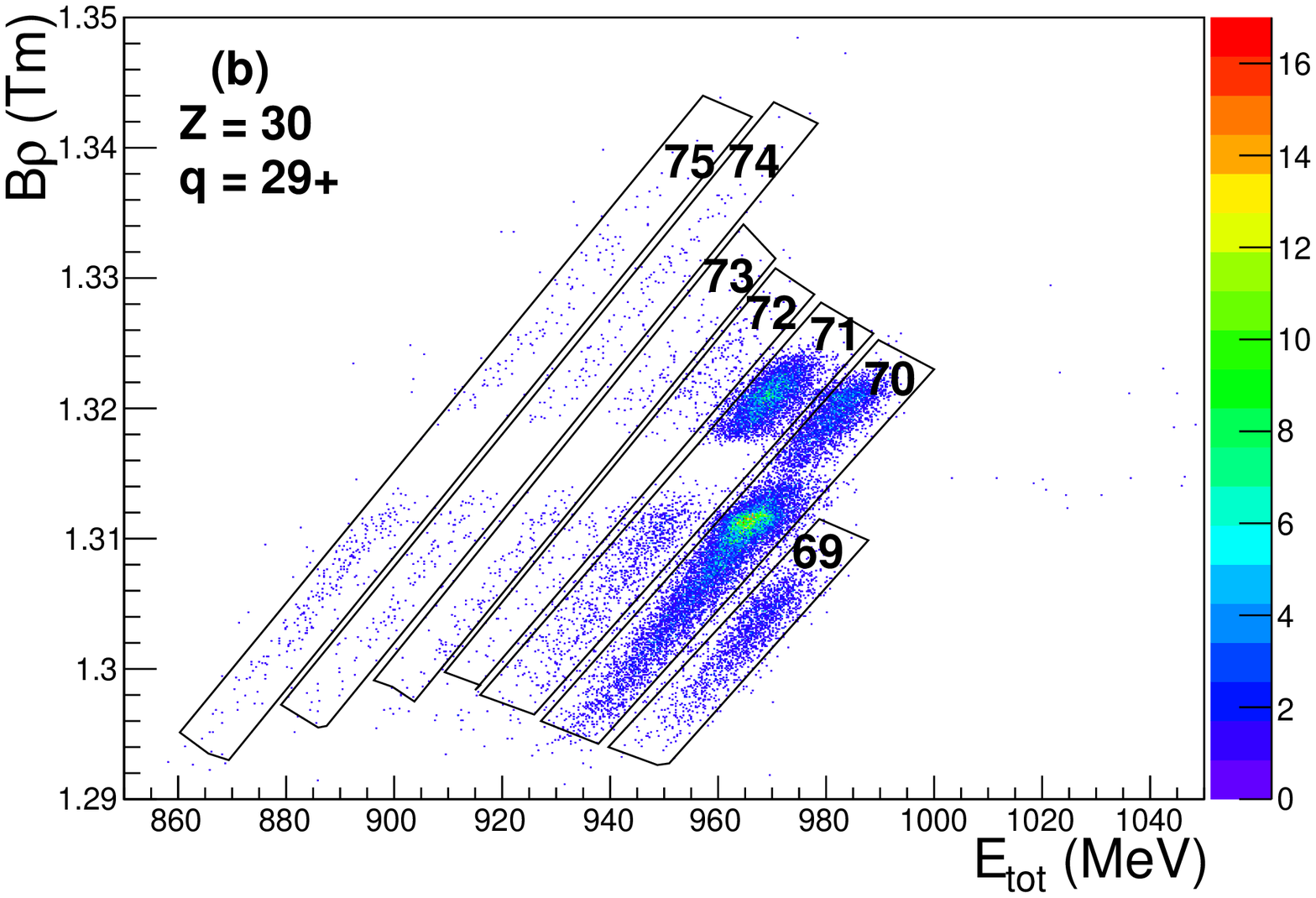}
\caption{ (Color online)   
Magnetic rigidity vs E$_{tot}$ correlation of ejectiles from  the 
$^{70}$Zn (15 MeV/nucleon) + $^{64}$Ni  reaction corresponding  
to the gates (drawn in Fig. 4) of Z=30, q=28 in (a),  
and Z=30, q=29 in (b). 
The gap in B$\rho$ is caused by the gate set to exclude 
$^{70}$Zn$^{29+}$ elastics, as explained in relation to Fig. 2,
and applied also to get Fig. 4.
Graphical contours are drawn on each band representing ejectiles of 
Zn$^{28+}$ with A=65--72 in (a), and Zn$^{29+}$ with A=69--75 in (b).
}
\label{fig05}
\end{figure}
Following the above discussion,  we present in Fig. 5 
the B${\rho}$ versus E$_{tot}$ correlation of events corresponding to 
Zn$^{28+}$ (Z=30, q=28) in panel (a), and Zn$^{29+}$ (Z=30, q=29) in panel (b).
These events were selected by the graphical contours of Z=30 and q=28, 29  shown in Fig. 4.
Adequate separation between the different mass bands of Zn ejectiles is achieved.
(One-dimensional projection of these bands gives a mass resolution of $\sim$0.6 A units.)
We note that the gap in the B$\rho$ range is due to the gate applied to exclude the 
elastic events corresponding to $^{70}$Zn$^{29+}$, as explained in relation to Fig. 2 
and correspondingly to Figs. 3 and 4.
In the B${\rho}$ versus E$_{tot}$ plot, the selection of the various
masses can be performed by setting the respective graphical cuts, as shown in 
Fig. 5a for A=65--72, and Fig. 5b for  A=69--75.
As we see in Fig. 5, the identification of neutron-rich isotopes with A$>$70 is clean, 
indicating that the goal of the present effort to produce and identify 
neutron-rich isotopes can be achieved.  
\begin{figure}[ht]                        
\centering
\includegraphics[width=0.45\textwidth,keepaspectratio=true]{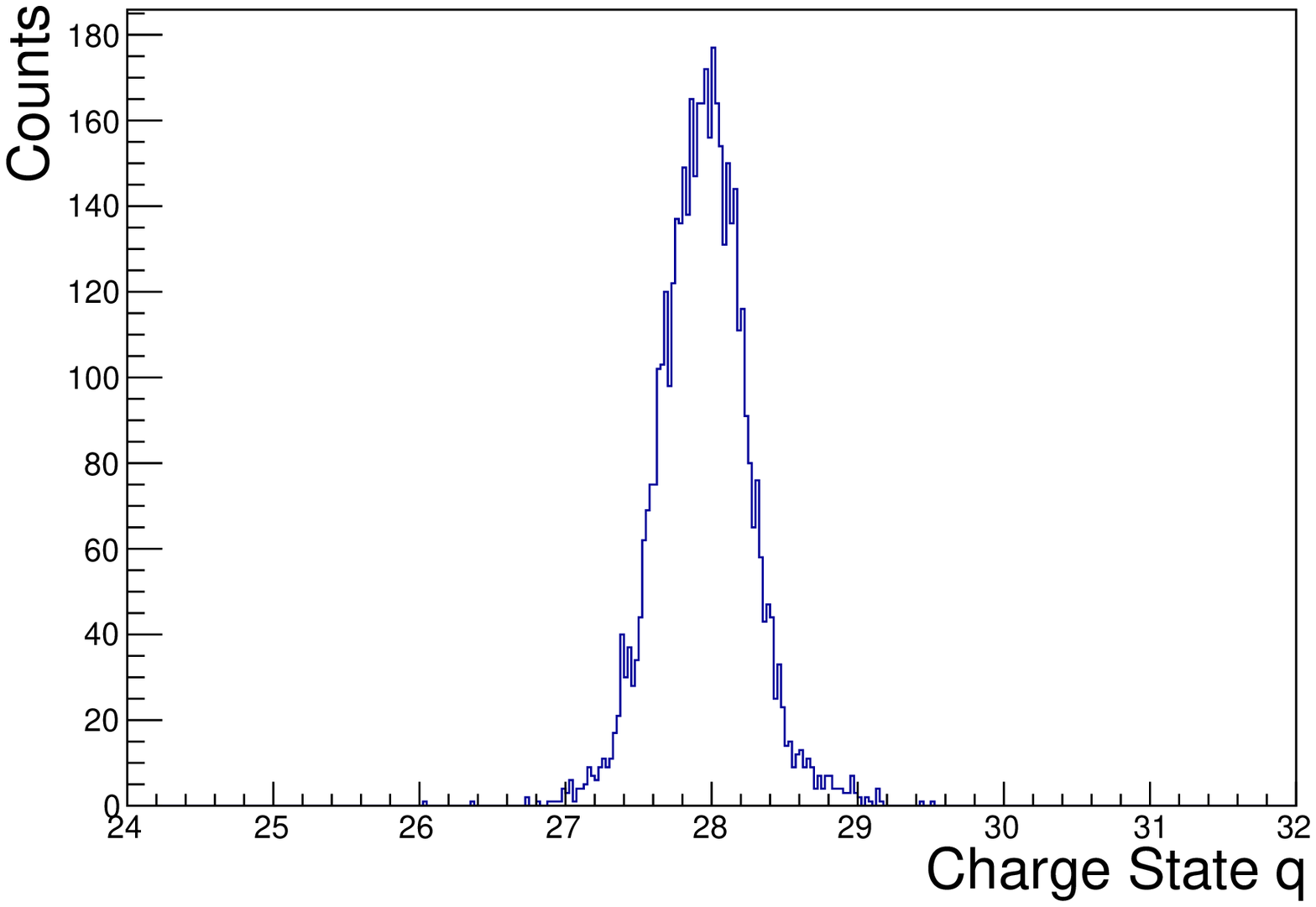}
\caption{ (Color online)   
Charge state spectrum obtained from the Z=30 gate of Fig. 4 and
the A=68 gate of the B${\rho}$ versus E$_{tot}$ correlation
indicated in  Fig. 5a (see text).
}
\label{fig06}
\end{figure}

At this point we note that since the magnetic rigidity is mainly determined
by the position  at the focal plane of the spectrometer and,   
only a small fraction of the energy is deposited in the gas, the relationship
of Eq. (5) is approximately maintained between the two primary measured quantities
 X$_{foc}$  and E$_{resid}$ and provides the basis of the mass identification procedure 
described in  \cite{Cappuzzello-2010}, in which absolute calibration of the 
involved quantities is not necessary.

Furthermore, we point out that for heavy ions, the q reconstruction is 
indispensable along with the Z reconstruction.
If the PID procedure was based only on the Z reconstruction, 
we could set gates on the Z variable to select  the atomic number and 
subsequently proceed with the mass identification as above.
However, for heavy ions, the B${\rho}$ versus E$_{tot}$ correlation
(or, equivalently, the X$_{foc}$ versus E$_{resid}$ correlation)  is not adequate 
to separate species characterized  by (nearly) the same ratio of $\sqrt{m}$/q,  but having  
different masses and different charge states.
One such  case is that  of  $^{68}$Zn$^{28+}$  and $^{73}$Zn$^{29+}$ ions which 
essentially  appear on the same band on the B${\rho}$ versus E$_{tot}$ plot for a 
contour of ``A=68'' 
($\sqrt{m}$/q = 0.2945 for $^{68}$Zn$^{28+}$ and $\sqrt{m}$/q = 0.2946  for $^{73}$Zn$^{29+}$).
This situation is demostrated in the charge state spectrum of Fig. 6, obtained 
by employing only the gate Z=30 of Fig. 4 (without q gate) and the contour A=68 
as in Fig. 5a. (The intense elastic peak $^{70}$Zn$^{29+}$ has been eliminated as 
in the previous figures.)
As expected, the spectrum has a main peak of q=28 representing $^{68}$Zn$^{28+}$ ions 
that correspond to the events inside the A=68 gate of Fig. 5a,
and a lower peak of q=29 representing $^{73}$Zn$^{29+}$ ions  
that correspond to the A=73 gate of Fig. 5b.

We wish to point out that the methodology presented in this section, 
concerning the Z, q, and A identification  of the ejectiles in one of the
Si detectors, has been implemented in the other active Si detectors of the FPD
and has been validated in all the runs of the present experimental data.


In closing, we commend that, regarding the velocity determination, 
it is important to carefully reconstruct the trajectory length of the ion and combine 
it with the TOF measurement.
Improvement in the TOF resolution for medium-mass heavy ions 
as previously mentioned, can be achieved by improving the cyclotron RF timing, 
and, furthermore, by using  a thin start detector at the location of the 
stripper foil at the entrance of the spectrometer. 
This detector will also provide precise entrance-angle information 
($\theta_{i}$,$\phi_{i}$) that will enhance the accuracy of the trajectory
reconstruction (especially for the vertical component) and, thus, further improve
the achieved mass and energy resolution \cite{Cunsolo-2002a,Cunsolo-2002b}.


\section{Discussion and Conclusions}  

In the present work,  we successfully  investigated the feasibility 
for identifying  medium-mass ejectiles in heavy ion reactions 
at 15 MeV/nucleon by using the large acceptance spectrometer 
MAGNEX at INFN-LNS.  
The developed approach, of course, can be applied to any large acceptance
raytracing spectrometer equipped with similar detectors.
In this work, we employed multinucleon transfer reactions of 
a $^{70}$Zn beam at 15 MeV/nucleon  interacting  with a $^{64}$Ni target.
The particle identification approach developed in \cite{Cappuzzello-2010}
for large acceptance magnetic spectrometers was extended with the measurement of the 
time-of-flight and the detailed reconstruction of the atomic number Z and the 
ionic charge q of the ejectiles.
The Z reconstruction is based on the energy loss, the residual energy
and the TOF measurements following appropriate absolute calibration.
The q reconstruction, furthermore requires the magnetic rigidity 
obtained from ion optical trajectory reconstruction.
For a given Z and q,  the mass identification is based on the correlation of
magnetic rigidity and total energy
that results in a separation  of particles (identified in Z  and q) according 
to their mass number A.
We note that, along the lines of the present work, the implementation of the TOF 
was employed in MAGNEX measurements within the context of the NUMEN project 
\cite{Agodi-2015,Cappuzzello-2018} 
involving ejectiles from $^{20}$Ne--induced reactions and resulted in a clean separation
of charge states of these lighter ions (see Fig. 4 of \cite{Cavallaro-2020a}).

The present work with the heavier $^{70}$Zn-like ejectiles at 15 MeV/nucleon 
indicates the demanding requirement of a detailed analysis approach for the  
measurements with a large acceptance spectrometer. 
With the procedures developed in this work,  a clear identification of the 
ejectiles and extraction of their distributions can be achieved.
With the trajectory reconstruction technique \cite{Cappuzzello-2011}, the energy of the 
identified ions can be obtained with a resolution of about 1/1000 (limited by the emittance
of the K800 cyclotron beams) that essentially enables charged particle spectroscopy, 
i.e., the direct evaluation of states populated in the explored reactions.

In this context, 
after proper identification of the ejectiles, we plan to obtain their 
energy spectra, angular distributions and production cross sections. 
Of primary interest  are the multineutron pickup products (e.g. up to $^{78-80}$Zn). 
Comparisons with reaction models are in line, extending our systematic studies of reaction 
mechanisms in this energy range \cite{Fountas-2014,Papageorgiou-2018,Palli-2021}.

The successful development of the identification approach for very neutron-rich isotopes
from these reactions in the energy of 15 MeV/nucleon
allows to extend these measurements at LNS with MAGNEX after the 
planned upgrade of the K800 cyclotron and the concurrent upgrade  of the MAGNEX focal 
plane detector system in the framework  of the NUMEN project \cite{Agodi-2021}.
Apart from the systematic study of the production cross sections of extremely neutron-rich
nuclides in these reactions,  it will be possible to study their structure  directly (via charged particle spectroscopy),  to the extent that the achievable high-energy resolution will allow for these medium-mass nuclei.
In addition, we envision coincident $\gamma$-ray or neutron-decay spectroscopy with appropriate detector arrays positioned around the target chamber 
(e.g. \cite{Lunardi-2007,Cavallaro-2016,Oliv-2018,Finoc-2020,Cappu-2021}),
in the same approach as in the PRISMA or VAMOS facilities.

In parallel to the above studies, the high-resolution energy and angle reconstruction possibility
allows novel investigations of the various nucleon transfer channels in the Fermi energy domain. 
Apart from the prevailing mechanism of sequential nucleon exchange 
\cite{Fountas-2014,Papageorgiou-2018,DIT,DITm,Veselsky-2011}, 
the transfer of nucleon pairs (up to 3, or more) can be investigated for the first time 
near and below the  Fermi energy and comparisons with recent theoretical developments 
(e.g. \cite{Agodi-2018}) can be performed.
Along this line, the possibility of direct transfer of clusters (d, $^{3}$He, t, $\alpha$)
in these collisions can be also explored and compared with appropriate theoretical models.


In closing, we envision the application of these developments and plans at upcoming RIB facilities
that can provide  
neutron-rich rare-isotope beams in the energy range  10--25 MeV/nucleon 
(e.g. \cite{GANIL,RISP,Tshoo-2013,Tshoo-2016,FRAISE}) enabling a rich agenda of  exciting
nuclear reaction and nuclear structure studies, in parallel to those at lower or 
higher (fragmentation) energy facilities.


\section{Acknowledgement}

We are grateful to the support staff of INFN--LNS for providing the primary beam. 
We are  thankful to  Dr Y.K. Kwon and Dr K. Tshoo for  enlighting discussions
on relevant plans with reaccelerated rare isotope beams at RISP.
This project has received funding from the European Research Council (ERC) under 
the European Union’s Horizon 2020 research and innovation programme 
(grant agreement No 714625). 
GS and SK ackowledge support from the Special Account for Research Grants of
the National and Kapodistrian  University of Athens. 
MV was supported by the  Czech Science Foundation (GACR) grant No. 21-24281S.






\end{document}